# Multi-Interface PLC / Wireless Network Simulation *

Yvonne-Anne Pignolet[1], Ilias Rinis[2], Dacfey Dzung[1], Abdulkadir Karaagac[3]

[1]ABB Corporate Research, Baden, [2]ETH Zurich, [3]EPF Lausanne, Switzerland

{yvonne-anne.pignolet, dacfey.dzung}@ch.abb.com,abdulkadir.karaagac@epfl.ch iliasr@gmail.com

*Abstract*—Many communication networks consist of legacy and new devices using heterogeneous technologies, such as copper wire, optical fiber, wireless and power line communication (PLC). Most network simulators, however, have been designed to work well with a single underlying link layer technology. Furthermore, there are hardly any suitable models for network simulators of PLC. In this paper we present extensions of the Contiki OS network simulator Cooja: A device may support multiple interfaces accessing multiple PLC segments or wireless channels and a simple PLC medium model is introduced describing packet loss probability as a function of distance. We test our approach to simulate a Smart Grid scenario of Ring Main Units equipped with PLC devices.

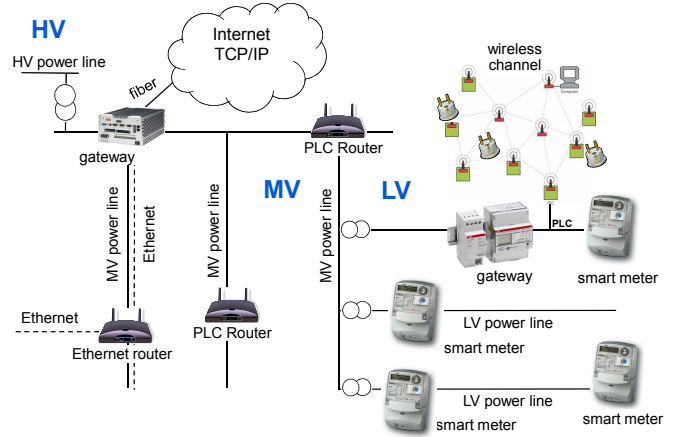

Fig. 1. Possible grid communication network using a combination of Ethernet, PLC and wireless technologies.

## I. INTRODUCTION

The Smart Grid vision predicts the existence of more than one available communication interface to offer redundancy and reliability as illustrated in Figure 1. This raises the challenge of combining them all into a single robust communication infrastructure to tackle the problems of Lossy and Low-Power networks (LLNs). As an example, one can imagine a network of devices that are capable of communicating over wireless connections, but also over PLC, or have two PLC interfaces to bypass an open electrical switch or a transformer. Thus networking stacks that take the characteristics of different communication technologies into account and can handle several one-hop connections between two devices are needed. Figure 2 shows an example of an IPv6 multi-interface stack.

The concurrent usage of a PLC and a RF interface on the MAC layer has been demonstrated in [7] on a customized sensor platform running Contiki. To enhance packet delivery ratio and throughput in a sensor network, two link selection algorithms are presented; one for low quality asymmetric links and one that quickly switches to the alternative link when conditions vary temporarily. However, the approach focuses only on the selection of the best link for a specific transmission, not on the selection of whole paths.

For the networking layer, the IETF Working Group Routing over Low power and Lossy networks (ROLL) proposed the IPv6 Routing Protocol for Low-power and Lossy Networks (RPL) [10]. RPL is a distance vector routing protocol designed to be communication medium agnostic and making decision based on objective functions and metrics that can depend on the underlying physical network. Recently communication stacks with RPL were introduced for wireless IEEE 802.15.4 networks [8], [4] and for powerline communication [3] for the operation systems Contiki and TinyOS.

**Our Contributions:** Since real deployments and verification in the field are costly, especially for PLC, simulation tools are necessary for pre-deployment testing. In this article we present network simulator extensions for the Contiki OS simulator Cooja such that devices can have several interfaces, e.g., a direct combination of wireless and PLC or other communication technologies is possible without requiring a gateway function. Cooja now features (a) a simple *MV PLC medium model*, simulating the effect of interference and attenuation in MV PLC on packet success rate, (b) each device can have *one or several interfaces* and (c) there can be *one or several media in the simulation setting* and each interface is attached to one of them (thus a device can have two interfaces accessing the same medium or different media).

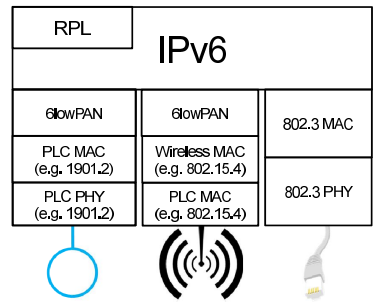

Fig. 2. Example communication stack using IPv6 and RPL with three interfaces. Each of the interfaces uses a different medium.

## II. MV PLC Medium Model

The definition of a realistic channel model is complex for PLC and there are no well-accepted models for PLC network level simulation. Apart from many other factors, in medium voltage networks the packet success rate is strongly dependent on the distance between the two nodes that communicate, hence we focus here on a simple approximation that assumes an exponential signal attenuation with distance. For other simulation models more targeted at low voltage use cases we refer to [1], [5], [9] and to [2] for a model ignoring distance-related attenuation.

Our model is based on the following assumptions

1) Attenuation for given cable and frequency is exponential w.r.t. distance $d$, i.e., $P_{RX} = P_{TX} \cdot e^{-\gamma \cdot d}$, where $P_{RX}$ is the received power, $P_{TX}$ is the tranmission power and $\gamma$ in [dB/m].
2) Packet success rate of a 50 bytes packet is 80% at 2km distance from sender, which corresponds to a bit error rate $BER = 5.6 \cdot 10^{-4}$.
3) $P_b = erfc(\sqrt{SNR/2})/2$ for BPSK.
4) Due interference effects the highest possible transmission success rate is at most $1 - 10^{-6}$, corresponding to $SNR_0 = 15.3$ dB.

Thus the following fading behavior is expected.

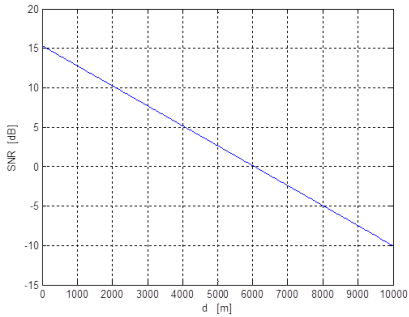

Fig. 3. SNR vs Distance for PLC Channel Model

Our assumptions lead to a bit error rate as depicted in Figure 4.

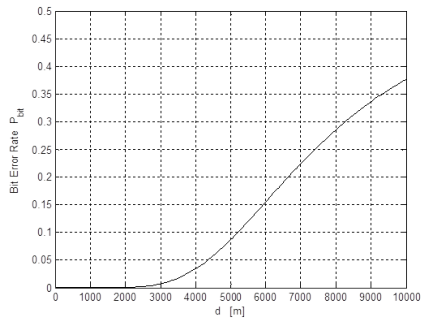

Fig. 4. BER vs Distance for PLC Channel Model

The packet success rate decreases with the number of bits $Nbits$ the packet contains, i.e., $P_{suc} = (1-P_b)^{Nbits}$. Figure 5 shows how the success rate behaves in this model for a packet of 50 bytes.

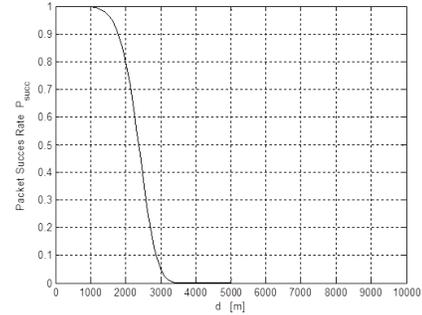

Fig. 5. Example success rate over distance for PLC (50 bytes)

Clearly this is a simplistic model that neglects many influences and which needs to be verified in field tests. However it serves as a first order approximation of distance-dependency for medium voltage PLC, so one can simulate relevant scenarios before deployment. Note that link delay is not relevant on MV PLC networks, because the signal propagation delays on electric wires over these distances is orders of magnitudes smaller than the upper networking layer processing delays.

In PLC, all devices that are physically connected by a powerline are in principle able to communicate with each other directly in a one-hop manner, although the achievable communication quality may vary. The transmission success rate of the MV PLC model uses the notion of distance to be the total distance among the shortest PLC path that connects the source with the destination.

## III. Cooja Network Simulator

The Contiki OS is accompanied by a network simulator called Cooja. It provides both a simulation engine and a user interface and enables the user to build networks, create and execute Contiki applications on nodes, observe log messages and control the network and the nodes with scripts.

Several radio communication models are implemented in Cooja, which specifies the connectivity, interference, transmission, receipt and all behavioral details of the medium. One of them is the Directed Graph Radio Medium (DGRM), which specifies for each link the transmission success rate.

**MVPLC Medium:** For the MV PLC model, we built a wrapper around DGRM, i.e., MV PLC medium configures DGRM to model an actual MV PLC grid. The user can specify a powerline graph connecting the interfaces of the devices, then the MV PLC medium computes the distance-dependent success probabilities used by the DGRM. To this end, it computes the connected components (disjoint subgraphs of nodes with paths among each other) of the powerline graph and adds a clique (complete graph) to the DGRM for each of the connected components. This property is illustrated in Figure 6.a) where four nodes are attached to a power line. The resulting communication graph is a complete graph (all nodes can communicate with all others), see Figure 6.b). We provide a simple MV PLC Configurator, depicted in Figure 6.c), where a user can specify a PLC graph and the underlying DGRM graph is computed automatically.

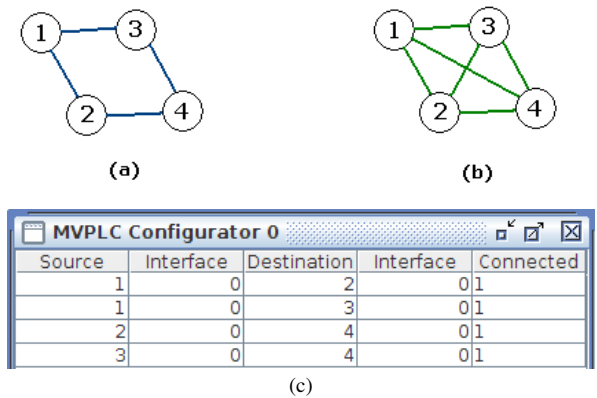

Fig. 6. A MV PLC ring a) and its DGRM graph model representation b) and the Cooja MV PLC configurator c).

The transmission success rate of the MV PLC model uses the total distance of the shortest PLC path that connects the source with the destination. For example, let us assume that in Figure 6.a) the shortest path between 1 and 4 is (1,2,4); hence the distance between 1 and 4 for the link is equal to the distance of (1,2) plus the distance of (2,3). With this distance the packet success rate is computed according to (1).

Simultaneous transmissions at distances where the packet success rate is above a certain threshold (e.g. 5%) are assumed to collide and are not received. In the implementation this is achieved by setting an interference range parameter.

**Multi-Interface Support:** Cooja was extended for more than one active communication media in the same simulation, enabling separate concurrent communication channels. Defining a single medium for a simulation is constraining in the existence of multiple interfaces and does not match all usage scenarios. To overcome this, each of the interfaces must be associated with a medium to model the communication channel of the interface. The medium class is modified such that one-to-one and one-to-many associations are possible: one medium for each interface, or one medium for multiple interfaces. Each medium operates independently.

**Smart Grid Scenario Simulation:** In order to realistically map a small MV PLC network topology to a network with devices employing more than one communication interfaces using our extensions, we consider a scenario with two nodes deployed at a primary substation and two separate ring lines with 7 nodes each deployed at Ring Main Units (RMUs) in the distribution network. Each of these nodes has two PLC couplers to be able to communicate on both lines attached, even if an electrical switch, breaker or transformer between them prevents direct communication among them. An example topology of the network is shown in Figure 7. The two nodes at the substation (node 1 and 9) are assumed to collect messages sent by the nodes at the ring main units. As a consequence the nodes have two PLC interfaces and use the MV PLC communication model described above. The rings are open between their third and forth nodes, i.e., the power in the grid is supplied radially. The distance between two neighboring ring main units is chosen uniformly at random in the range of [200m, 2000m], as typical for Europe's distribution networks. The green edges represent MV PLC lines; the red "x" between nodes 1 and 8 denotes the switch that later will be opened to simulate a failure.

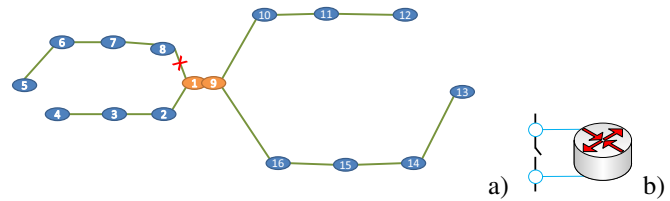

Fig. 7. a) Example MV PLC network topology with two (open) rings. b) Each RMU is equipped with a PLC router attached to the line with two couplers.

## IV. CONCLUSION

In this article we present network simulator extensions for multiple lossy communication interfaces. We extended the Cooja network simulator such that it handle multiple communication interfaces and provided simple medium voltage PLC communication channel model. Such a simulator is especially useful for Smart Grid scenarios using several communication technologies and/or PLC devices with two couplers.

REFERENCES

[1] L. Ben Saad, C. Chauvenet, B. Tourancheau, et al. Simulation of the RPL Routing Protocol for IPv6 Sensor Networks: Two Cases Studies. In *Sensor Technologies and Applications SENSORCOMM*, 2011.
[2] G. Bumiller. Power-line Physical Layer Emulator for Protocol Development. In *International Symposium on Power-Line Communications and its Applications*, 2004.
[3] C. Chauvenet, B. Tourancheau, D. Genon-Catalot, P. Goudet, and M. Pouillot. A Communication Stack over PLC for Multi Physical Layer IPv6 Networking. In *Smart Grid Communications*, pages 250–255, 2010.
[4] J. Ko, J. Eriksson, N. Tsiftes, S. Dawson-Haggerty, A. Terzis, A. Dunkels, and D. Culler. ContikiRPL and TinyRPL: Happy Together. In *IPSN 2011*, 2011.
[5] E. Malacasa and G. Morabito. Characterization of PLC Communication Channel: a Networking Perspective. In *Workshop on Power Line Communication (WSPLC)*, 2009.
[6] Y.-A. Pignolet, I. Rinis, D. Dzung, and A. Karaagac. Multi-Interface Extensions for PLC / Wireless Simulator. In *Sixth Workshop on Power Line Communication*, 2012.
[7] Y. Sun, S. Bhadra, S. Choi, M. Fu, and X. Lu. Achieving Robust Sensor Networks through Close Coordination between Narrow-band Power Line and Low-Power Wireless Communications. In *Embedded Networked Sensor Systems*, pages 373–374, 2011.
[8] N. Tsiftes, J. Eriksson, and A. Dunkels. Low-Power Wireless IPv6 Routing with ContikiRPL. In *IPSN*, 2010.
[9] F. Versolatto and A. Tonello. Analysis of the PLC Channel Statistics using a Bottom-up Random Simulator. In *International Symposium on Power Line Communications and Its Applications (ISPLC)*, pages 236–241. IEEE, 2010.
[10] T. Winter, P. Thubert, A. Brandt, T. H. Clausen, J. W. Hui, R. Kelsey, P. Levis, K. Pister, R. Struik, and J. Vasseur. RPL: IPv6 Routing Protocol for Low power and Lossy Networks. *IETF RFC 6550*, 2012.